\documentclass{article}

\usepackage{microtype}
\usepackage{graphicx}
\usepackage{subcaption}
\usepackage{booktabs}
\usepackage{amsmath}
\usepackage{multirow}
\usepackage{siunitx}
\usepackage{tabularx}
\usepackage{makecell}
\usepackage{hyperref}

\usepackage[preprint]{icml2026}

\usepackage{amssymb}
\usepackage{mathtools}
\usepackage{amsthm}
\usepackage[capitalize,noabbrev]{cleveref}

\theoremstyle{plain}

\theoremstyle{definition}

\theoremstyle{remark}

\usepackage[textsize=tiny]{todonotes}

\icmltitlerunning{ShapeMark: Robust and Diversity-Preserving Watermarking for Diffusion Models}

\begin{document}

\twocolumn[
  \icmltitle{ShapeMark: Robust and Diversity-Preserving Watermarking for Diffusion Models}



  \icmlsetsymbol{equal}{*}

  \begin{icmlauthorlist}
      \icmlauthor{Yuqi Qian}{inst1,inst2}
      \icmlauthor{Yun Cao}{inst1,inst2}
      \icmlauthor{Haocheng Fu}{inst1,inst2}
      \icmlauthor{Meiyang Lv}{inst1,inst2}
      \icmlauthor{Meineng Zhu}{inst3}
    \end{icmlauthorlist}

    \icmlaffiliation{inst1}{Institute of Information Engineering, Chinese Academy of Sciences, Beijing, China.}
    \icmlaffiliation{inst2}{School of Cyber Security, University of Chinese Academy of Sciences, Beijing, China.}
    \icmlaffiliation{inst3}{School of Cybersecurity, University of International Relations, Beijing, China.}

    \icmlcorrespondingauthor{Yun Cao}{caoyun@iie.ac.cn}

  \icmlkeywords{Machine Learning, ICML}

  \vskip 0.3in
]



\printAffiliationsAndNotice{}  

\begin{abstract}
Diffusion models have made substantial advances in recent years, enabling high-quality image synthesis; however, the widespread dissemination and reuse of their outputs have introduced new challenges in intellectual property protection and content provenance.
Image watermarking offers a solution to these challenges, and recent work has increasingly explored Noise-as-Watermark (NaW) approaches that integrate watermarking directly into the diffusion process.
However, existing NaW methods fail to balance robustness and diversity.
We attribute this weakness to value encoding, which encodes watermark bits into individual sampled values. It is extremely fragile in practical application scenarios.
To address this, we encode watermark bits into the structured noise pattern, so that the watermark is preserved even when individual values are perturbed. 
To further ensure generation diversity, we introduce a dedicated randomization design that reshuffles the positions of noise elements without changing their values, preventing the watermark from inducing fixed noise patterns or spatial locations.
Extensive experiments demonstrate that our method achieves state-of-the-art robustness while maintaining high generation quality across a wide range of lossy scenarios.
\end{abstract}

\section{Introduction}
Recent advances in diffusion models~\cite{ho2020denoising,song2020denoising,rombach2022high,peebles2023scalable} have enabled large-scale, high-fidelity image synthesis, significantly accelerating the adoption of generative models in real-world content creation. 

As diffusion-based generative models are increasingly adopted in practical AIGC pipelines, the need to protect the copyright of generated images and to enable the detection of unauthorized reuse has become increasingly pressing, especially as generated content is widely disseminated and routinely subjected to lossy post-processing in real-world scenarios.
At the same time, any provenance mechanism must preserve the core advantages of diffusion models, including visual fidelity and generation diversity. These requirements call for watermarking techniques that can provide reliable ownership signals while remaining tightly aligned with the generative process.

Among existing solutions, Noise-as-Watermark (NaW) methods have emerged as a particularly appealing paradigm for watermarking diffusion models. In NaW, watermark information is embedded into the initial noise used for image generation and subsequently recovered by inverting the diffusion process. Since the watermark is introduced at the noise level before the generative process begins, NaW methods avoid direct modification of either the generated images or the underlying model parameters, allowing watermarking to be naturally integrated into the diffusion pipeline and readily applied across different models and deployment settings.

\begin{figure}[t]
    \centering
    \includegraphics[
        width=\columnwidth,
        trim=20 0 0 20,
        clip
    ]{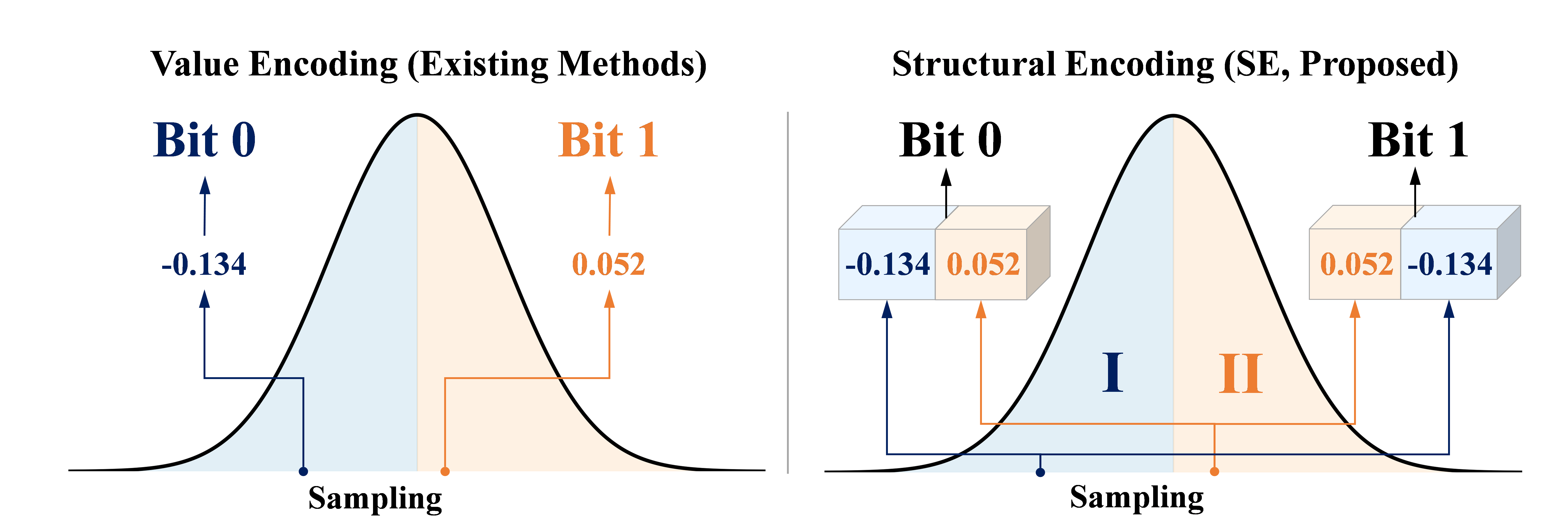}
    \caption{Comparison between existing Noise-as-Watermark methods and the proposed method.}
    \label{fig:introduction}
\end{figure}

However, existing NaW methods struggle to simultaneously achieve robust watermark detectability and high generation diversity.
First, most existing methods rely on \textit{value encoding}, in which watermark information is mapped to individual sampled values, for example by constraining their sign, magnitude, or sampling range. Under this design, reliable watermark detection implicitly depends on accurate recovery of individual noise values. However, in lossy scenarios involving post-processing and imperfect diffusion inversion, such precise value recovery is inherently unreliable. As illustrated in Fig.~\ref{fig:introduction}, most watermark-carrying elements are close to the decision boundary (e.g., around zero),  so that even small perturbations to the noise can flip their extracted states and lead to incorrect decoding.
Second, many NaW approaches strengthen robustness by repeatedly embedding watermark information during the noise sampling process. We observe that this operation leads to a noticeable reduction in the diversity of the generated watermarked images and simultaneously makes the watermark artifacts more readily perceptible.

To address these challenges, we propose \textbf{ShapeMark}, a noise-as-watermark method designed to jointly achieve strong robustness and high generation diversity.
Our main contributions are summarized as follows:
\begin{itemize}
\item \textbf{Structural Encoding (SE).} We propose a robust watermark encoding scheme that extends noise-as-watermark from individual value encoding to structural encoding across multiple noise elements. 
SE embeds watermark bits by encoding them into shapes formed by groups of noise elements sampled from separable regions of the noise distribution. 
Consequently, verification depends on recovering group-level permutations that can be reliably reconstructed despite perturbations.

\item \textbf{Payload-Debiasing Structural Randomization (PDSR).} We identify that the diversity degradation in existing watermarking pipelines stems from the fact that they introduces fixed and payload dependent patterns into the sampled noise. To address this issue, we propose PDSR, a lightweight structural randomization strategy that reshuffles the positions of the noise elements after embedding while strictly keeping their values unchanged. By decoupling the payload identity from a fixed spatial pattern, PDSR enables the same watermark to correspond to multiple noise realizations, effectively preserving image diversity.

\item Extensive experiments demonstrate that ShapeMark achieves strong robustness across a wide range of lossy processing operations; even under high-intensity distortions, it consistently maintains true positive rates exceeding 99\% at a fixed low false positive rate, while maintaining high overall generation diversity.
\end{itemize}

\section{Related Work}
\subsection{Diffusion Models and Diffusion Inversion}

Diffusion models generate data by learning to reverse a gradual noise corruption process~\cite{ho2020denoising,song2020denoising,rombach2022high}. Given a data sample $x_0 \sim q(x)$, the forward diffusion process progressively adds Gaussian noise over $T$ steps according to a predefined variance schedule, resulting in noisy variables $x_t$. This process admits a closed-form expression:
\begin{equation}
x_t = \sqrt{\bar{\alpha}_t}\,x_0 + \sqrt{1-\bar{\alpha}_t}\,\epsilon, 
\quad \epsilon \sim \mathcal{N}(0,I),
\label{eq:forward_diffusion}
\end{equation}
where $\bar{\alpha}_t = \prod_{i=1}^t (1-\beta_i)$.

In diffusion models, image generation proceeds by transforming an initial noise sample $x_t \sim \mathcal{N}(0,I)$ into a clean image $x_0$.
Under Denoising Diffusion Implicit Models (DDIM)~\cite{song2020denoising}, the clean image at step $t$ can be estimated as
\begin{equation}
\hat{x}_0^{\,t} =
\frac{x_t - \sqrt{1-\bar{\alpha}_t}\,\epsilon_\theta(x_t)}
{\sqrt{\bar{\alpha}_t}},
\end{equation}
which defines a deterministic denoising trajectory.

The deterministic formulation of DDIM further enables an approximate inversion process, which maps a generated image back to its corresponding noise latent.
Although this inversion is approximate, prior work has shown that the recovered noise is often sufficiently close to that used during generation, making it suitable for latent-space watermark verification.

\subsection{Image Watermarking for Diffusion Models}
With the rapid adoption of diffusion models in image generation, watermarking techniques tailored to diffusion-based frameworks have attracted increasing attention. Existing diffusion watermarking methods can be broadly categorized into three groups.
\textbf{Post-processing approaches} embed watermarks directly into generated images~\cite{ingemar2008digital,zhang2019robust,jia2021mbrs,ma2022towards}. Although easy to deploy, these methods often exhibit limited robustness under lossy compression or geometric distortions and may introduce perceptible visual artifacts.
\textbf{Model-based approaches} inject watermark information by fine-tuning diffusion models or their associated components~\cite{fernandez2023stable,min2024watermark,kingma2013auto}. While achieving good imperceptibility, they require additional training and suffer from reduced scalability across different models and deployment settings.
In contrast, \textbf{Noise-as-Watermark (NaW) approaches} embed watermark information into the initial noise of the diffusion process and recover it through diffusion inversion. \cite{wen2023tree,ci2024ringid} introduce robust pattern features in the frequency domain, but inevitably cause deviations from the original noise distribution. Other works~\cite{huang2024robin,zhang2024attack} focus on zero-bit watermarking for source verification via optimization-based embedding, resulting in extremely limited payload capacity. More recent approaches~\cite{yang2024gaussian,gunn2024undetectable} enable high-capacity embedding through distribution-preserving sampling, while T2SMark~\cite{yang2025t2smark} further improves robustness by encoding bits in statistically reliable regions via tail-truncated sampling.
Despite methodological differences, existing NaW approaches predominantly rely on value encoding, where watermark information is directly tied to individual noise values or their sampled ranges. Consequently, they are vulnerable to perturbations that disrupt the precise value correspondence, limiting robustness in practical scenarios.

\begin{figure*}[t]
    \centering
    \includegraphics[
        width=0.8\textwidth,
        trim=20 30 0 30,
        clip
    ]{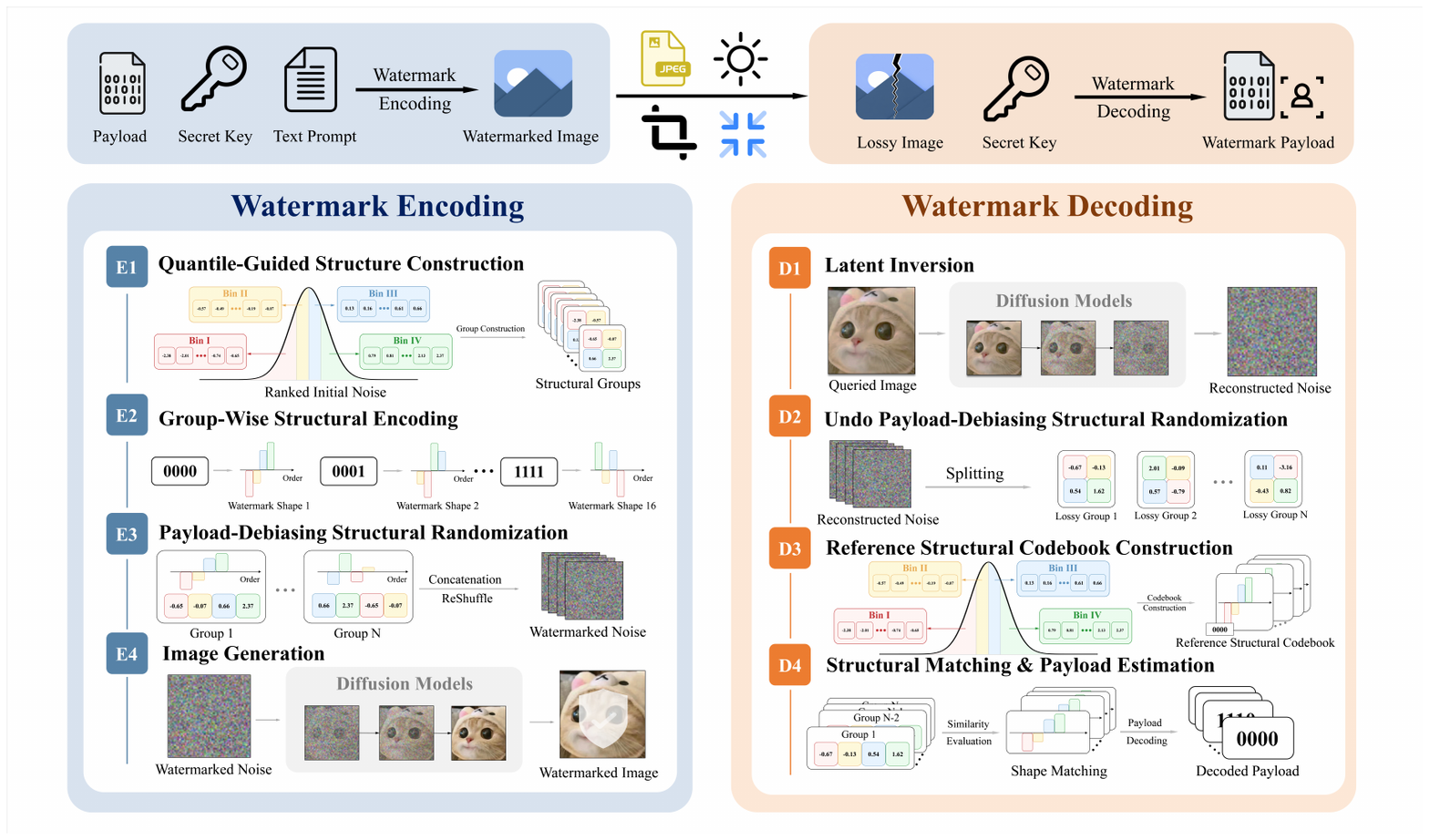}

    \caption{Overall pipeline of ShapeMark. The figure illustrates structural-level watermark encoding in the diffusion noise latent and structure-based watermark decoding via diffusion inversion and reference codebook matching. For clarity of visualization, each block is shown with a block size of one (i.e., a single noise element per block), whereas the actual method supports larger block sizes in practice.}
    \label{Overview}
\end{figure*}

\section{Method}
\subsection{Threat Model and Application Scenarios}
\paragraph{Threat model and applications.}
We study keyed watermarking for diffusion-generated images, where a secret key specifies the operations related to the watermarking process. An attacker knows the algorithm but not the key, and can only access the final image, applying common post-processing distortions while preserving visual plausibility. Verification operates from a queried image under weak conditioning via diffusion inversion, and decides watermark presence by a calibrated threshold, optionally decoding payload bits. ShapeMark targets multi-tenant generative services for identity binding, provenance verification, and traceability (e.g., user/session metadata) in platform-side screening and attribution workflows.

\subsection{Overview of ShapeMark}
As illustrated in Fig.~\ref{Overview}, the proposed ShapeMark consists of an encoding and a corresponding decoding processes.

\paragraph{Watermark Encoding.}
During watermark encoding, the initial noise latent is sampled and processed by Structural Encoding (SE), which embeds the watermark payload by rearrange the noise elements' positions. Then we apply Payload-Debiasing Structural Randomization (PDSR) to process the noise in order to remove its fixed structural pattern.
The resulting latent noise is subsequently fed into the diffusion model, together with the text prompt, to generate the final watermarked image.

\paragraph{Watermark Decoding.}
During watermark decoding, the queried image undergoes standard diffusion inversion and undoing PDSR to obtain the initial noise containing watermark information. The verification party reconstructs the noise without rearrangement based on the shared information, and calculates the shape similarity between the two to extract the watermark information.

\subsection{Structural Encoding (SE)}
\label{sec:SE}
SE embeds watermark payloads into the initial noise latent by rearranging the noise elements' position, without altering their values.
The core idea is to construct a standardized index template guided by quantile information, and then embed watermark information by using the payload to control the shape of structural units within this template.

SE consists of two stages:
(i) construct a separability-guided template that maps \emph{elements}
\(\rightarrow\) \emph{quantile bins} \(\rightarrow\) \emph{blocks} \(\rightarrow\) \emph{groups};
(ii) encode each payload chunk by a group-wise block permutation selected from a
permutation codebook.

\paragraph{Stage I: construct a standardized index template.}

\textbf{(1) Rank elements and form magnitude-quantile bins.}
We first compute the magnitude ranking of all coordinates:
\begin{equation}
\pi = \mathrm{argsort}(|z|) \in S_D,
\quad |z_{\pi(1)}| \le \cdots \le |z_{\pi(D)}|.
\label{eq:SE_rank}
\end{equation}
We then split the ranked indices into \(Q\) equal-sized bins by rank.
Let \(\ell \triangleq D/Q\) and define the \(q\)-th quantile bin as
\begin{equation}
B_q = \{\pi((q-1)\ell+1), \dots, \pi(q\ell)\}, \quad q=1,\dots,Q.
\label{eq:SE_bins}
\end{equation}
Intuitively, indices in larger-$q$ bins correspond to noise elements with larger magnitudes and are therefore statistically more distinguishable from those in smaller-$q$ bins.

\textbf{(2) Construct blocks and groups.}
Given the quantile bins $\{B_q\}_{q=1}^Q$ from Step (1), we next build the two higher-level
structural units used for encoding: \emph{blocks} (within-bin bundles) and \emph{groups}
(cross-bin alignments).

\textbf{Blocks (within each bin).}
For each bin $B_q$ (of size $\ell=D/Q$), we use the secret key $\kappa$ to generate a
pseudo-random but reproducible ordering of its indices, and denote
the resulting ordered list by $\widetilde{B}_q$.
We then bundle indices from the same bin into blocks by taking consecutive chunks of length $b$.
Let $T \triangleq \ell/b$ be the number of blocks per bin, and define the $t$-th block in bin $q$ as
\begin{equation}
I_{q,t} =
\big(\widetilde{B}_q[(t-1)b+1], \dots, \widetilde{B}_q[tb]\big),
\quad t=1,\dots,T.
\label{eq:SE_blocks}
\end{equation}
Each $I_{q,t}$ is an index tuple identifying a \emph{block} of $b$ coordinates.
Importantly, all elements in a block come from the same quantile bin $B_q$; hence,
blocks originating from different bins correspond to statistically separable magnitude regions.
The key $\kappa$ controls \emph{which} indices from $B_q$ are bundled together by determining
$\widetilde{B}_q$, while leaving all values unchanged.

\textbf{Groups (aligning one block per bin).}
A \emph{group} collects one block from each bin, yielding $Q$ blocks per group.
In our canonical construction, we form groups by aligning blocks with the same block index $g$
across bins:
\begin{equation}
G \triangleq T,\quad
\mathcal{G}_g = (I_{1,g}, I_{2,g}, \dots, I_{Q,g}), \quad g=1,\dots,G.
\label{eq:SE_groups}
\end{equation}
In the default setting, the group size is $s=Q$ (one block per bin).
Although Eq.~\eqref{eq:SE_groups} uses a fixed alignment rule, the resulting group membership
is still \emph{key-controlled} because each constituent block $I_{q,g}$ is determined by the
key-shuffled ordering $\widetilde{B}_q$ in Eq.~\eqref{eq:SE_blocks}.
By construction, blocks are disjoint, and thus groups are disjoint as well.

\paragraph{Stage II: payload-conditioned within-group block rearrangement.}
With the template \(\{\mathcal{G}_g\}_{g=1}^G\) fixed, we embed payload bits by
permuting \emph{which quantile-derived block occupies which group position}.

\textbf{Codebook.}
Let \(k\) be the number of payload bits carried per group.
We choose a permutation codebook \(C \subseteq S_Q\) with \(|C|=2^k\), together with
a fixed encoder \(\mathrm{Enc}:\{0,1\}^k \rightarrow C\).
Given a payload bitstring \(m\), we split it into \(G\) chunks
\(m_1,\dots,m_G \in \{0,1\}^k\) and map each chunk to a codeword permutation
\(\sigma_g = \mathrm{Enc}(m_g)\in C\).

\textbf{Embedding by within-group block permutation.}
Let \(z^{(e)}\) denote the SE-encoded latent, initialized as \(z^{(e)} \leftarrow z\).
For each group \(g\) and each group position \(j\in[Q]\), we reassign the block values
according to \(\sigma_g\):
\begin{equation}
z^{(e)}[I_{j,g}] = z[I_{\sigma_g(j),g}],
\quad \forall g\in[G],\ \forall j\in[Q].
\label{eq:SE_embed}
\end{equation}
Equivalently, each group defines \(Q\) destination ``slots'' \((I_{1,g},\dots,I_{Q,g})\),
and \(\sigma_g\) specifies which source block (and thus which quantile-derived values)
is moved into each slot.
Because different groups are disjoint, Eq.~\eqref{eq:SE_embed} can be applied to all
groups in parallel.

\paragraph{Why SE is robust.}
SE enhances robustness by encoding watermark information in \emph{relative structural
relationships} rather than in individual noise values.
Specifically, the payload is carried by the permutation pattern of quantile-derived blocks
within each group, which is substantially more stable under practical distortions than
absolute element values.
While inversion error or post-processing noise may perturb individual coordinates, such
perturbations are unlikely to consistently alter the underlying within-group block ordering.

Moreover, each block aggregates multiple noise elements, so decoding operates at the
block level instead of the element level.
This aggregation has an implicit averaging effect, reducing the influence of lossy
distortions on watermark recovery.
Importantly, the entire embedding remains value distribution-preserving: SE only permutes
the coordinates of an existing Gaussian sample without modifying any value.

\subsection{Payload-Debiasing Structural Randomization (PDSR)}
\label{sec:pdsr}

\paragraph{Motivation.}
SE encodes bits via a deterministic within-group block permutation configuration.
If the same payload $m$ is reused across many generations under a fixed identity key $\kappa$,
the resulting SE-aligned configurations may exhibit \emph{payload-dependent spatial regularities}:
even though SE is value-preserving, the blocks originating from particular quantile bins can
tend to recur in similar spatial regions after encoding.
Such regularities may (i) reduce perceptual diversity across repeated generations with the same payload,
and (ii) increase detectability of watermark-induced patterns by correlating spatial artifacts across images.

PDSR addresses this issue by applying an additional \emph{payload-agnostic} global mixing step that is
(i) value-preserving, (ii) exactly invertible, and crucially (iii) \emph{one-time randomized per image}
via a public nonce, so that the same payload can correspond to diverse noise realizations.

\paragraph{Definition and invertibility.}
Let $z^{(e)}$ denote the SE-encoded latent and let $\{\mathcal{I}_n\}_{n=1}^{N}$ be the ordered list of all blocks ($N=D/b$).
We derive a nonce-conditioned key and sample a pseudo-random permutation over blocks:
\begin{equation}
\begin{aligned}
\kappa_{\mathrm{pdsr}}
&= \mathrm{KDF}\!\big(\kappa, r), \\
\tau(r)
&= \mathrm{PRP}_{\kappa_{\mathrm{pdsr}}}(N) \in S_N .
\end{aligned}
\end{equation}

which depends only on $(\kappa,r)$ and is independent of the payload.
We then permute block \emph{locations} globally:
\begin{equation}
z^{(p)}[\mathcal{I}_n] \;=\; z^{(e)}\big[\mathcal{I}_{\tau(r)(n)}\big],
\qquad \forall n\in[N].
\label{eq:pdsr}
\end{equation}
Since PDSR is a pure permutation, it is exactly invertible given $\tau(r)$:
\begin{equation}
z^{(e)}[\mathcal{I}_n] \;=\; z^{(p)}\big[\mathcal{I}_{\tau(r)^{-1}(n)}\big],
\qquad \forall n\in[N].
\label{eq:inv-pdsr}
\end{equation}

\paragraph{Why it helps.}
PDSR breaks the direct linkage between (i) the payload-dependent within-group configuration induced by SE
and (ii) fixed spatial placement in the latent.
The per-image nonce $r$ ensures that even under a fixed identity key $\kappa$ and a repeated payload $m$,
the global placement permutation $\tau(r)$ changes from image to image, yielding \emph{diverse noise realizations}
and preventing repeated generations from sharing consistent spatial artifacts.
Because PDSR is payload-agnostic, value-preserving, and exactly invertible, it improves diversity and
imperceptibility without weakening the structural cues that SE relies on for decoding (since the verifier
can deterministically undo the same permutation using $(\kappa, r)$).

\subsection{Watermark Detection and Payload Decoding}
\label{sec:detection}

Given a queried image $I$, the verifier is provided with the identity key $\kappa$,
the SE/PDSR hyperparameters $(Q,b,s)$, and the shared permutation codebook
$\mathcal{C}\subseteq S_s$ with encoder/decoder $\mathrm{Enc}:\{0,1\}^k\!\to\!\mathcal{C}$
and $\mathrm{Dec}=\mathrm{Enc}^{-1}$.
The verifier aims to (i) determine whether $I$ contains a valid watermark under $\kappa$, and
(ii) optionally recover the embedded payload.

\paragraph{Stage I: recover the SE-aligned watermarked latent.}
This step outputs an estimate $\hat{z}^{(e)}$ that corresponds to the SE-encoded latent,
up to diffusion inversion error and channel distortions. 

\emph{Diffusion inversion to the initial noise latent.}
We first invert the queried image into the diffusion initial-noise space using a standard inversion operator (e.g., DDIM inversion):
\begin{equation}
\hat{z}^{(p)} \;=\; \mathcal{D}^{\dagger}_{\theta}(I;\tilde{c}),
\end{equation}
where $\tilde{c}$ denotes the (possibly unknown or mismatched) conditioning used at verification time.
The recovered latent $\hat{z}^{(p)}$ approximates the generation-time PDSR-randomized latent $z^{(p)}$.

\emph{Reconstruct the structural template and undo PDSR.}
Using $\kappa$, the verifier regenerates the canonical Gaussian latent $z$ and reconstructs the same bins/blocks/groups
as in SE, yielding the group index tuples
\begin{equation}
\{\mathcal{G}_g\}_{g=1}^{G}, \quad
\mathcal{G}_g = (\mathcal{I}_{1,g},\ldots,\mathcal{I}_{s,g}),\;\;\; \mathcal{I}_{j,g}\in [D]^b,
\end{equation}
and an ordered list of all blocks $\{\mathcal{I}_n\}_{n=1}^{N}$ with $N=D/b$, consistent with Sec.~\ref{sec:SE}.
(In our default setting, $s=Q$.)

\emph{Undoing PDSR via a public nonce.}
In addition to $\kappa$, the verifier obtains the public per-image nonce $r$ associated with $I$
(e.g., from a provenance record or generation log).
We derive a PDSR key to reconstruct the corresponding block permutation $\tau(r)$, which is used to undo PDSR.

The resulting $\hat{z}^{(e)}$ is aligned to the SE embedding space and serves as the input to decoding.

\paragraph{Stage II: codebook matching for watermark detection and payload decoding.}
This stage compares the recovered latent structure in $\hat{z}^{(e)}$ against the shared codebook $\mathcal{C}$,
using the canonical latent $z$ regenerated from $\kappa$ as the reference.

\emph{Group-wise matching score against the shared codebook.}
For each group $\mathcal{G}_g=(\mathcal{I}_{1,g},\ldots,\mathcal{I}_{s,g})$, we read the observed blocks from $\hat{z}^{(e)}$:
\begin{equation}
\hat{b}_{j,g} \;=\; \hat{z}^{(e)}[\mathcal{I}_{j,g}] \in \mathbb{R}^{b},
\qquad j=1,\ldots,s,
\end{equation}
and the corresponding reference blocks from the canonical latent $z$:
\begin{equation}
b^{\mathrm{ref}}_{j,g} \;=\; z[\mathcal{I}_{j,g}] \in \mathbb{R}^{b},
\qquad j=1,\ldots,s.
\end{equation}
For each candidate codeword permutation $\sigma\in\mathcal{C}$, we define the group-wise matching score
\begin{equation}
d_g(\sigma) \;=\; \sum_{j=1}^{s}\left\|\hat{b}_{j,g} - b^{\mathrm{ref}}_{\sigma(j),g}\right\|_2^2.
\label{eq:group-score}
\end{equation}

\emph{Payload recovery.}
We decode the group codeword by selecting the best-matching permutation in the codebook:
\begin{equation}
\hat{\sigma}_g \;=\; \arg\min_{\sigma\in\mathcal{C}} d_g(\sigma),
\qquad
\hat{m}_g \;=\; \mathrm{Dec}(\hat{\sigma}_g)\in\{0,1\}^k.
\label{eq:decode}
\end{equation}
Concatenating $\{\hat{m}_g\}_{g=1}^{G}$ yields the recovered payload $\hat{m}$.

\emph{Watermark existence test.}
For identity verification, the verifier typically checks a \emph{claimed} key/identity, in which case the expected group codewords
$\{\sigma_g^\star\}_{g=1}^G$ (or equivalently, the expected payload) are known under $\kappa$.
We quantify evidence of watermark presence via a per-group margin between the target codeword and its best competitor:
\begin{equation}
d_g^{\mathrm{comp}} \;=\; \min_{\sigma\in\mathcal{C}\setminus\{\sigma_g^\star\}} d_g(\sigma),
\qquad
d_g^{\mathrm{tgt}} \;=\; d_g(\sigma_g^\star),
\end{equation}
\begin{equation}
m_g \;=\; \max\!\left(0,\; \frac{d_g^{\mathrm{comp}} - d_g^{\mathrm{tgt}}}{d_g^{\mathrm{comp}}+\varepsilon}\right),
\end{equation}
where $\varepsilon>0$ is a small constant.
The global detection statistic aggregates margins across groups:
\begin{equation}
S \;=\; \frac{1}{G}\sum_{g=1}^{G} m_g.
\label{eq:global-score}
\end{equation}
We declare watermark presence if $S\ge \tau_{\mathrm{det}}$, where $\tau_{\mathrm{det}}$ is calibrated to achieve a desired false positive rate
(e.g., using non-watermarked images and/or verification under incorrect keys).

\paragraph{Complexity.}
Since $s$ and $|\mathcal{C}|$ are small in practice (e.g., $s=4$, $|\mathcal{C}|=16$), Stage II decoding is efficient:
$\mathcal{O}(G\cdot |\mathcal{C}|\cdot s\cdot b)$.

\begin{figure*}[h]
    \centering
    \includegraphics[
        width=0.8\textwidth,
        trim=20 0 0 20,
        clip
    ]{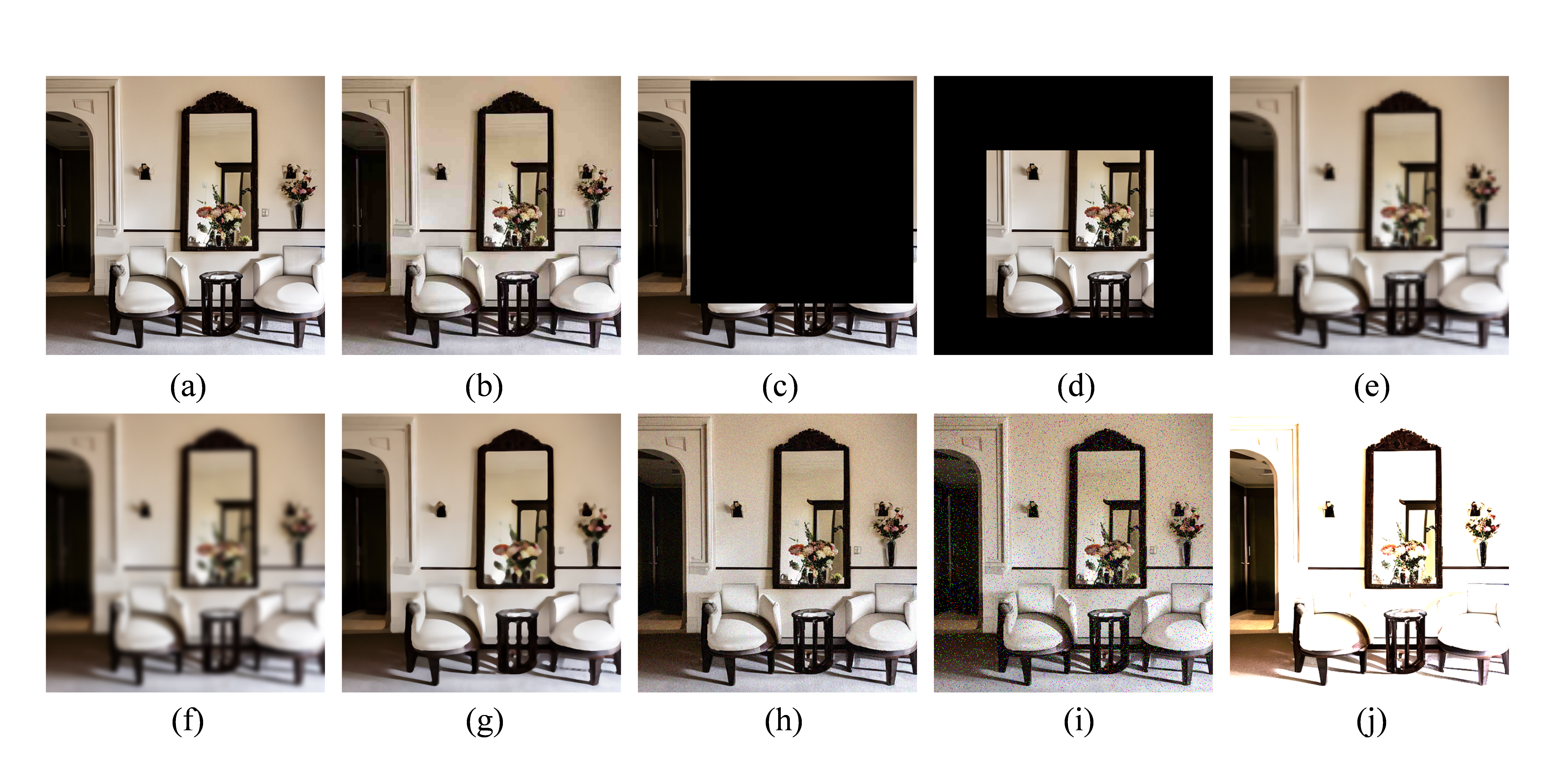}
    \caption{The watermarked image is attacked by different distortions. (a) Watermarked image. (b) JPEG, QF = 25. (c) 80\% area Random Drop. (d) 60\% area Random Crop.  (e) 25\% Resize and restore. (f) Gaussian Blur, $r$ = 4. (g) Median Blur, $k$ = 7. (h) Gaussian Noise, $\mu$ = 0, $\sigma$ = 0.05. (i) Salt and Pepper Noise, $p$ = 0.05. (j) Brightness, factor = 6.}
    \label{lossy}
\end{figure*}

\begin{table*}[ht]
\centering
\caption{Quantitative comparison of watermarking methods in terms of robustness (TPR at FPR = $10^{-6}$), traceability (per-bit accuracy), generation diversity (LPIPS), and visual quality (CLIP score and FID). Results are reported for both clean and attacked settings. Standard errors and two-sample t-statistics are provided for CLIP score and FID to assess the impact of watermarking on image quality.}
\small
\setlength{\tabcolsep}{6pt}
\begin{tabular}{lccccc}
\toprule
Method & TPR@FPR=$10^{-6}$ & Bit Acc. & Diversity$\uparrow$ & CLIP Score ($t\downarrow$) & FID ($t\downarrow$) \\
\midrule
SD v2.1~\cite{rombach2022high}
& -- & -- 
& 0.7091 
& $0.3076_{\scriptsize \pm 0.0002}$ {\scriptsize (--)} 
& $56.4198_{\scriptsize \pm 0.26}$ {\scriptsize (--)} \\
\midrule
dwtDct~\cite{ingemar2008digital}
& 0.921 / 0.181
& 0.8115 / 0.5925
& -- 
& $0.3072_{\scriptsize \pm 0.0002}$ {\scriptsize (4.47)} 
& $55.2946_{\scriptsize \pm 0.25}$ {\scriptsize (9.86)} \\

dwtDctSvd~\cite{ingemar2008digital}
& 0.998 / 0.479 
& 0.9909 / 0.7040
& -- 
& $0.3073_{\scriptsize \pm 0.0002}$ {\scriptsize (3.35)} 
& $54.9795_{\scriptsize \pm 0.23}$ {\scriptsize (13.12)} \\

RivaGAN~\cite{zhang2019robust}
& 0.916 / 0.429
& 0.9769 / 0.8728
& -- 
& $0.3073_{\scriptsize \pm 0.0002}$ {\scriptsize (3.35)} 
& $55.1793_{\scriptsize \pm 0.27}$ {\scriptsize (10.47)} \\

StableSig~\cite{fernandez2023stable}
& 0.999 / 0.426
& 0.9912 / 0.7647
& 0.6982
& $0.3078_{\scriptsize \pm 0.0002}$ {\scriptsize (2.83)} 
& $54.8369_{\scriptsize \pm 0.19}$ {\scriptsize (15.54)} \\

TRW~\cite{wen2023tree}
& \textbf{1.000} / 0.898 
& -- / -- 
& 0.6941 
& $0.3073_{\scriptsize \pm 0.0001}$ {\scriptsize (4.24)} 
& $59.1735_{\scriptsize \pm 0.20}$ {\scriptsize (26.55)} \\

GS~\cite{yang2024gaussian}
& \textbf{1.000} / 0.993
& \textbf{1.0000} / 0.9755
& 0.6322 
& $0.3080_{\scriptsize \pm 0.0006}$ {\scriptsize (2.00)} 
& $58.8084_{\scriptsize \pm 0.65}$ {\scriptsize (10.79)} \\

PRCW~\cite{gunn2024undetectable}
& \textbf{1.000} / 0.580
& 0.9086 / 0.7763
& \underline{0.7258}
& $0.3075_{\scriptsize \pm 0.0002}$ {\textbf{\scriptsize (1.18)} }
& \textbf{$57.1114_{\scriptsize \pm 0.24}$} \textbf{{\scriptsize (6.18)}} \\

T2SMark~\cite{yang2025t2smark}
& \textbf{1.000} / \underline{0.998}
& \textbf{1.0000} / \underline{0.9859}
& 0.7032 
& $0.3077_{\scriptsize \pm 0.0002}$ \underline{{\scriptsize (1.23)}}
& $57.7566_{\scriptsize \pm 0.41}$ {\scriptsize (8.71)} \\

ShapeMark
& \textbf{1.000} / \textbf{0.999}
& \textbf{1.0000} / \textbf{0.9870}
& \textbf{0.7338}
& $0.3073_{\scriptsize \pm 0.0004}$ {\scriptsize (2.12)}
& $57.2958_{\scriptsize \pm 0.32}$ {\underline{\scriptsize (6.72)}} \\

\bottomrule
\end{tabular}
\label{tab:mainexp}
\end{table*}

\begin{table*}[t]
\centering
\caption{Ablation study evaluating the impact of key structural components in ShapeMark. Variants selectively remove separability-guided structural construction and payload-debiasing structural randomization while keeping the watermark capacity fixed. Results are reported in terms of robustness, generation diversity, and visual quality.}
\small
\setlength{\tabcolsep}{6pt}
\begin{tabular*}{\textwidth}{@{\extracolsep{\fill}}lcccc}
\toprule
Method & Bit Acc. & Diversity$\uparrow$ & CLIP Score ($t\downarrow$) & FID ($t\downarrow$) \\
\midrule
w/o SE, w/o PDSR
& 1.000/0.9374 & 0.6981 
& $0.3083 \pm 0.0003$ (6.1394) 
& $58.1347 \pm 0.28$ (15.0202) \\

w/ SE, w/o PDSR
& 1.000/0.9600
& 0.7049
& $0.3063 \pm 0.0006$ (11.4018) 
& $57.9574 \pm 0.36$ (14.0072) \\

w/o SE, w/ PDSR
& 1.000 / 0.9369
& 0.7303
& $0.3074 \pm 0.0005$ (1.1744) 
& $57.2765 \pm 0.19$ (8.4128) \\

ShapeMark
& 1.000 / 0.9870
& 0.7338
& $0.3073 \pm 0.0004$ (2.1213) 
& $57.2958 \pm 0.32$ (6.7186) \\
\bottomrule
\end{tabular*}
\label{tab:ablation}
\end{table*}

\section{Experiments}
\subsection{Experimental Settings}
\label{sec:exp-settings}

\paragraph{Implementation Details.}
Our method uses Stable Diffusion v2.1~\cite{rombach2022high} as the generative backbone. Unless otherwise specified, we keep the inference configuration fixed across methods: latent resolution $4\times 64\times 64$, classifier-free guidance scale $7.5$, 50 denoising steps, and output resolution $512\times 512$.
We use the default ShapeMark setting with $Q=4$ bins, block size $b=64$ latent elements, and group size $s=4$ blocks (one block per bin). Each group encodes $k=4$ bits, resulting in a total payload of 256 bits per image in our main setting.
For watermark verification, we recover an estimate of the initial noise via standard DDIM inversion. To simulate a weakly-conditioned setting where the original generation prompt is unavailable, we perform inversion using an empty prompt with 10 inversion steps and guidance scale set to 1. We apply the same inversion protocol to all baselines that require latent inversion, ensuring a consistent and challenging evaluation environment. All methods are implemented in PyTorch 2.6.0, and all experiments are conducted on a single NVIDIA RTX L40 GPU.

\paragraph{Baselines.}
We compare ShapeMark against representative methods from three paradigms:
(i) post-processing watermarking: dwtDct, dwtDctSvd~\cite{ingemar2008digital}, RivaGAN~\cite{zhang2019robust},
(ii) model fine-tuning watermarking (Stable Signature~\cite{fernandez2023stable}),
and (iii) noise-as-watermark methods (Tree-Ring~\cite{wen2023tree}, Gaussian Shading~\cite{yang2024gaussian}, PRC-Watermark~\cite{gunn2024undetectable}, and T2SMark~\cite{yang2025t2smark}).
For a capacity-matched comparison, methods that support high-capacity payloads are configured to embed 256 bits per image; RivaGAN and Stable Signature follow their official implementations (32 bits and 48 bits, respectively).

\paragraph{Datasets and Evaluation Metrics.}
We evaluate the performance of the method on MS-COCO 2017~\cite{lin2014microsoft} and the Stable-Diffusion-Prompts (SDP)~\cite{stable_diffusion_prompts_gustavosta} dataset.
For robustness and traceability, we sample 500 prompts from the SDP training split and generate 500 watermarked images per method. Each image is further subjected to nine lossy distortions (Shown in Fig.~\ref{lossy}) before detection and payload recovery.
We report (i) detection performance as TPR at a fixed FPR of $10^{-6}$, and (ii) traceability as payload recovery accuracy; attacked performance is aggregated over the nine distortions.
For methods that modify the generative process (excluding post-processing baselines), we measure generation diversity using LPIPS~\cite{zhang2018unreasonable}: we generate 10 images per prompt for 1,000 COCO validation prompts (with a fixed key and payload) and average LPIPS over all intra-prompt image pairs.
We assess visual quality using CLIP~\cite{radford2021learning} score and FID~\cite{heusel2017gans} on 1,000 COCO validation prompts, and report mean and standard error over multiple trials. Statistical testing details (including the two-sample $t$-test protocol) and implementation specifics are provided in Appendix~\ref{app:ttest}.

\subsection{Main Results}
\label{sec:main-results}

Table~\ref{tab:mainexp} compares robustness, traceability, diversity, and image quality across all methods. ShapeMark consistently performs best on the two objectives most critical for practical watermarking: it achieves a TPR of \textbf{0.999} at $\mathrm{FPR}=10^{-6}$ and a per-bit recovery accuracy of \textbf{0.9870} under distortions, indicating reliable detection and payload extraction in the presence of post-processing and inversion noise. Importantly, this robustness does not come at the expense of generative behavior: ShapeMark attains the \textbf{highest} diversity among all methods (LPIPS \textbf{0.7338}) while maintaining competitive image quality (CLIP $0.3073 \pm 0.0004$, FID $57.2958 \pm 0.32$), supporting our claim that structural encoding with payload-debiasing randomization mitigates payload-induced bias and preserves generation quality.

\begin{figure*}[ht]
    \centering
    \includegraphics[
        width=\textwidth,
        trim=20 0 0 20,
        clip
    ]{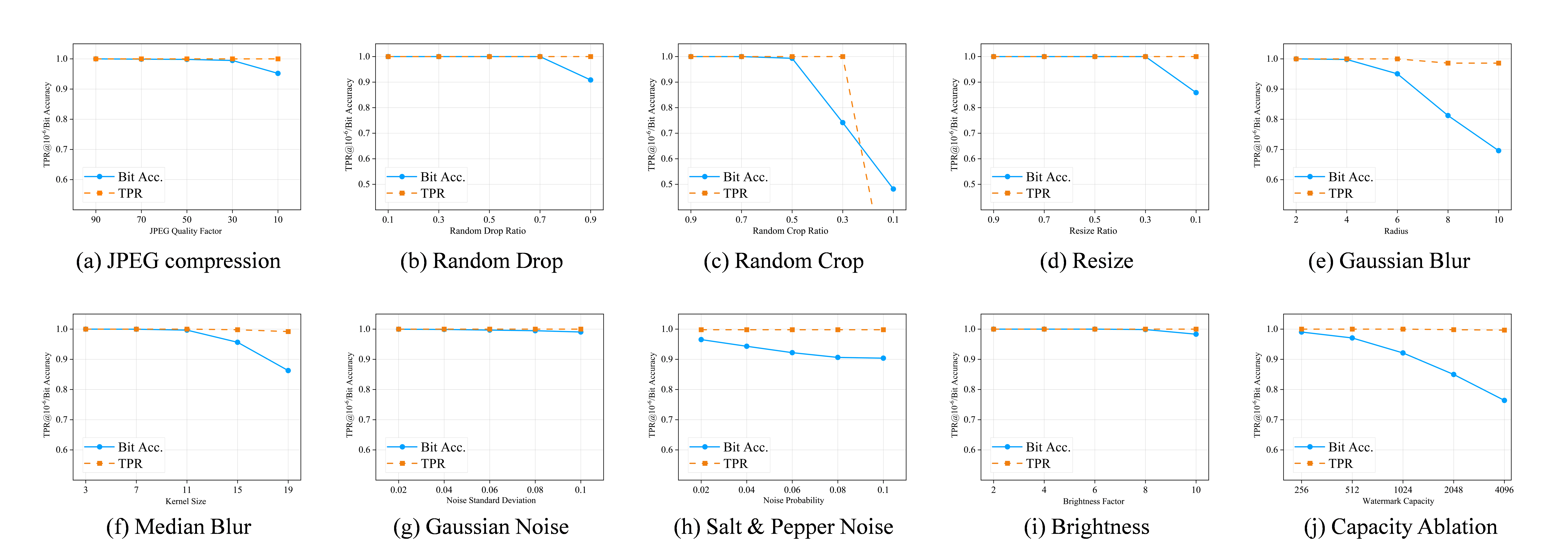}
    \caption{Ablation Studies.}
    \label{Ablation}
\end{figure*}

\subsection{Ablation Study}
\label{sec:ablation}

We ablate the key design choices and decoding hyperparameters of ShapeMark while keeping the remaining settings identical to the main experiments.

\paragraph{Effect of SE \& PDSR.}
We evaluate two component-level variants: (i) \emph{w/o separability-guided construction}, where we form blocks by sequentially grouping consecutive flattened latent entries and build groups by concatenation (all sizes and payload capacity unchanged); and (ii) \emph{w/o PDSR}, where we remove the  payload-agnostic mixing and rely only on the first-stage group-wise structural encoding. Table~\ref{tab:ablation} shows that removing the separability-guided construction substantially degrades robustness under attacks, confirming that forming groups from distinct quantile regions is critical for reliable extraction after distortions. In contrast, removing PDSR only mildly affects extraction accuracy but leads to a noticeable deterioration in perceptual behavior (e.g., diversity/imperceptibility), indicating that the second stage primarily mitigates payload-induced bias without being the main driver of robustness. Together, these results support the complementary roles of SE (robustness) and PDSR (quality/diversity preservation).

\textbf{Noise strength and distortion severity.} To further assess robustness, we evaluate watermark performance under a wide range of distortion types and severity levels, covering all lossy operations considered in our threat model. The results are summarized in Fig.~\ref{Ablation}. Across most evaluated scenarios, even under substantial distortions, ShapeMark consistently maintains reliable watermark detection performance.

In particular, ShapeMark exhibits strong robustness to additive Gaussian noise. Even at a relatively high noise level of $\sigma = 0.1$, ShapeMark achieves over 99\% payload recovery accuracy. This result is noteworthy given that prior noise-as-watermark methods have reported pronounced performance degradation under Gaussian perturbations. We attribute this robustness to ShapeMark’s structure encoding: while Gaussian noise perturbs individual latent values, the within-group block ordering remains recoverable, enabling reliable recovery even under strong noise.

\textbf{Effect of guidance scale and inversion steps.}
Finally, we vary verification-related hyperparameters while keeping the watermark configuration fixed. As reported in Table~\ref{tab:guidance_inversion}, stronger guidance during detection slightly reduces bit accuracy under attacks, likely because richer content increases inversion difficulty. Increasing the number of DDIM inversion steps yields a modest and consistent improvement, reflecting more accurate recovery of the initial latent. Notably, ShapeMark already achieves reliable extraction with a small number of inversion steps, suggesting that the method remains effective under weakly conditioned and low-cost verification settings.

\textbf{Effect of Watermark Capacity.} Finally, we investigate the effect of watermark capacity by adjusting the block size, which directly controls the overall payload capacity. In this experiment, we focus exclusively on watermark extraction performance under lossy conditions, evaluating robustness in the presence of distortions.
As expected, increasing the watermark capacity leads to a gradual degradation in detection and payload recovery performance. Notably, ShapeMark maintains strong robustness at high capacities: even at a payload size of 2048 bits, the proposed method still achieves approximately 85\% per-bit recovery accuracy under attacked settings. These results demonstrate the scalability of the proposed structural encoding mechanism and its ability to support high-capacity watermarking while retaining meaningful robustness.

\begin{table}[ht]
\centering
\caption{Effect of guidance scale during generation and the number of DDIM inversion steps during extraction on per-bit watermark recovery accuracy under attacked conditions.}
\label{tab:guidance_inversion}
\small
\setlength{\tabcolsep}{6pt}
\begin{tabular}{c ccccc}
\toprule
\multirow{2}{*}{\shortstack{Inversion\\Steps}}
& \multicolumn{5}{c}{Guidance Scales} \\
\cmidrule(lr){2-6}
& 4 & 6 & 8 & 10 & 12 \\
\midrule
5   & 0.9855 & 0.9828 & 0.9803 & 0.9778 & 0.9751 \\
10  & 0.9858 & 0.9832 & 0.9807 & 0.9785 & 0.9760 \\
25  & 0.9858 & 0.9832 & 0.9806 & 0.9783 & 0.9757 \\
50  & 0.9857 & 0.9831 & 0.9804 & 0.9781 & 0.9755 \\
100 & 0.9856 & 0.9830 & 0.9803 & 0.9779 & 0.9752 \\
\bottomrule
\end{tabular}
\end{table}

\section{Conclusion}
\label{sec:conclusion}

ShapeMark introduces a structure-level noise watermarking for diffusion models that achieves robust and generation diverse. We construct separability-guided blocks and groups from noise order statistics and encode payloads via group-wise block permutations (SE), making the watermark resilient to inversion error and common post-generation distortions. We further apply a payload-agnostic structural randomization (PDSR) to remove payload-induced spatial bias, preserving sampling diversity and visual fidelity without modifying model weights. Extensive experimental results support ShapeMark as a practical primitive for identity binding, provenance verification, and traceability in diffusion services.

\section*{Impact Statement}
This paper presents work whose goal is to advance the field of Machine Learning by improving the practicality of watermarking for generative image models. In particular, our method aims to embed watermarks while preserving visual quality and sample diversity, which can make watermarking more deployable in real-world content generation pipelines.

\textbf{Potential benefits.} Reliable and imperceptible watermarking can support content provenance and accountability for AI-generated imagery, helping platforms, researchers, and downstream users identify synthetic content, attribute model outputs, and mitigate harms associated with misinformation, fraud, and unauthorized redistribution. It may also assist creators and rights-holders in tracking and managing the use of generated media, and facilitate auditing and policy compliance in settings where disclosure of synthetic content is required.

\textbf{Potential risks and misuse.} Watermarking techniques can be dual-use. More robust and less perceptible watermarks may be misused for covert tagging, surveillance-like tracking, or restrictive DRM practices without users’ knowledge or consent. Additionally, as watermarking becomes more capable, adversarial dynamics may intensify (e.g., attempts to remove, spoof, or falsely attribute watermarks), which could be used to evade detection or to frame benign content as originating from a particular model or source.

\textbf{Mitigations and responsible use.} We encourage deployments that emphasize transparency and consent (clear disclosure when watermarking is applied), governance over who can embed and verify watermarks, and careful consideration of privacy and user autonomy. In high-stakes applications, watermarking should be combined with complementary measures (e.g., access control, cryptographic provenance, and platform-level policies) rather than relied upon as a sole mechanism. We also recommend evaluating systems against removal and spoofing attempts and documenting limitations to reduce over-reliance on watermark signals.


\bibliography{example_paper}

\begin{thebibliography}{23}
\providecommand{\natexlab}[1]{#1}
\providecommand{\url}[1]{\texttt{#1}}
\expandafter\ifx\csname urlstyle\endcsname\relax
  \providecommand{\doi}[1]{doi: #1}\else
  \providecommand{\doi}{doi: \begingroup \urlstyle{rm}\Url}\fi

\bibitem[Ci et~al.(2024)Ci, Yang, Song, and Shou]{ci2024ringid}
Ci, H., Yang, P., Song, Y., and Shou, M.~Z.
\newblock Ringid: Rethinking tree-ring watermarking for enhanced multi-key
  identification.
\newblock In \emph{European Conference on Computer Vision}, pp.\  338--354.
  Springer, 2024.

\bibitem[Fernandez et~al.(2023)Fernandez, Couairon, J{\'e}gou, Douze, and
  Furon]{fernandez2023stable}
Fernandez, P., Couairon, G., J{\'e}gou, H., Douze, M., and Furon, T.
\newblock The stable signature: Rooting watermarks in latent diffusion models.
\newblock In \emph{Proceedings of the IEEE/CVF International Conference on
  Computer Vision}, pp.\  22466--22477, 2023.

\bibitem[Gunn et~al.(2024)Gunn, Zhao, and Song]{gunn2024undetectable}
Gunn, S., Zhao, X., and Song, D.
\newblock An undetectable watermark for generative image models.
\newblock \emph{arXiv preprint arXiv:2410.07369}, 2024.

\bibitem[{Gustavosta}(2022)]{stable_diffusion_prompts_gustavosta}
{Gustavosta}.
\newblock Stable diffusion prompts dataset.
\newblock
  \url{https://huggingface.co/datasets/Gustavosta/Stable-Diffusion-Prompts},
  2022.
\newblock Accessed: 2026-01-23.

\bibitem[Heusel et~al.(2017)Heusel, Ramsauer, Unterthiner, Nessler, and
  Hochreiter]{heusel2017gans}
Heusel, M., Ramsauer, H., Unterthiner, T., Nessler, B., and Hochreiter, S.
\newblock Gans trained by a two time-scale update rule converge to a local nash
  equilibrium.
\newblock \emph{Advances in neural information processing systems}, 30, 2017.

\bibitem[Ho et~al.(2020)Ho, Jain, and Abbeel]{ho2020denoising}
Ho, J., Jain, A., and Abbeel, P.
\newblock Denoising diffusion probabilistic models.
\newblock \emph{Advances in neural information processing systems},
  33:\penalty0 6840--6851, 2020.

\bibitem[Huang et~al.(2024)Huang, Wu, and Wang]{huang2024robin}
Huang, H., Wu, Y., and Wang, Q.
\newblock Robin: Robust and invisible watermarks for diffusion models with
  adversarial optimization.
\newblock \emph{Advances in Neural Information Processing Systems},
  37:\penalty0 3937--3963, 2024.

\bibitem[Ingemar et~al.(2008)Ingemar, Matthew, Jeffrey, Jessica, and
  Ton]{ingemar2008digital}
Ingemar, C., Matthew, M., Jeffrey, B., Jessica, F., and Ton, K.
\newblock Digital watermarking and steganography, 2008.

\bibitem[Jia et~al.(2021)Jia, Fang, and Zhang]{jia2021mbrs}
Jia, Z., Fang, H., and Zhang, W.
\newblock Mbrs: Enhancing robustness of dnn-based watermarking by mini-batch of
  real and simulated jpeg compression.
\newblock In \emph{Proceedings of the 29th ACM international conference on
  multimedia}, pp.\  41--49, 2021.

\bibitem[Kingma \& Welling(2013)Kingma and Welling]{kingma2013auto}
Kingma, D.~P. and Welling, M.
\newblock Auto-encoding variational bayes.
\newblock \emph{arXiv preprint arXiv:1312.6114}, 2013.

\bibitem[Lin et~al.(2014)Lin, Maire, Belongie, Hays, Perona, Ramanan,
  Doll{\'a}r, and Zitnick]{lin2014microsoft}
Lin, T.-Y., Maire, M., Belongie, S., Hays, J., Perona, P., Ramanan, D.,
  Doll{\'a}r, P., and Zitnick, C.~L.
\newblock Microsoft coco: Common objects in context.
\newblock In \emph{European conference on computer vision}, pp.\  740--755.
  Springer, 2014.

\bibitem[Ma et~al.(2022)Ma, Guo, Hou, Yang, Li, Jia, and Xie]{ma2022towards}
Ma, R., Guo, M., Hou, Y., Yang, F., Li, Y., Jia, H., and Xie, X.
\newblock Towards blind watermarking: Combining invertible and non-invertible
  mechanisms.
\newblock In \emph{Proceedings of the 30th ACM International Conference on
  Multimedia}, pp.\  1532--1542, 2022.

\bibitem[Min et~al.(2024)Min, Li, Chen, and Cheng]{min2024watermark}
Min, R., Li, S., Chen, H., and Cheng, M.
\newblock A watermark-conditioned diffusion model for ip protection.
\newblock In \emph{European Conference on Computer Vision}, pp.\  104--120.
  Springer, 2024.

\bibitem[Peebles \& Xie(2023)Peebles and Xie]{peebles2023scalable}
Peebles, W. and Xie, S.
\newblock Scalable diffusion models with transformers.
\newblock In \emph{Proceedings of the IEEE/CVF international conference on
  computer vision}, pp.\  4195--4205, 2023.

\bibitem[Radford et~al.(2021)Radford, Kim, Hallacy, Ramesh, Goh, Agarwal,
  Sastry, Askell, Mishkin, Clark, et~al.]{radford2021learning}
Radford, A., Kim, J.~W., Hallacy, C., Ramesh, A., Goh, G., Agarwal, S., Sastry,
  G., Askell, A., Mishkin, P., Clark, J., et~al.
\newblock Learning transferable visual models from natural language
  supervision.
\newblock In \emph{International conference on machine learning}, pp.\
  8748--8763. PmLR, 2021.

\bibitem[Rombach et~al.(2022)Rombach, Blattmann, Lorenz, Esser, and
  Ommer]{rombach2022high}
Rombach, R., Blattmann, A., Lorenz, D., Esser, P., and Ommer, B.
\newblock High-resolution image synthesis with latent diffusion models.
\newblock In \emph{Proceedings of the IEEE/CVF conference on computer vision
  and pattern recognition}, pp.\  10684--10695, 2022.

\bibitem[Song et~al.(2020)Song, Meng, and Ermon]{song2020denoising}
Song, J., Meng, C., and Ermon, S.
\newblock Denoising diffusion implicit models.
\newblock \emph{arXiv preprint arXiv:2010.02502}, 2020.

\bibitem[Wen et~al.(2023)Wen, Kirchenbauer, Geiping, and
  Goldstein]{wen2023tree}
Wen, Y., Kirchenbauer, J., Geiping, J., and Goldstein, T.
\newblock Tree-ring watermarks: Fingerprints for diffusion images that are
  invisible and robust.
\newblock \emph{arXiv preprint arXiv:2305.20030}, 2023.

\bibitem[Yang et~al.(2025)Yang, Fang, Zhang, Yu, and Chen]{yang2025t2smark}
Yang, J., Fang, H., Zhang, W., Yu, N., and Chen, K.
\newblock T2smark: Balancing robustness and diversity in noise-as-watermark for
  diffusion models.
\newblock \emph{arXiv preprint arXiv:2510.22366}, 2025.

\bibitem[Yang et~al.(2024)Yang, Zeng, Chen, Fang, Zhang, and
  Yu]{yang2024gaussian}
Yang, Z., Zeng, K., Chen, K., Fang, H., Zhang, W., and Yu, N.
\newblock Gaussian shading: Provable performance-lossless image watermarking
  for diffusion models.
\newblock In \emph{Proceedings of the IEEE/CVF Conference on Computer Vision
  and Pattern Recognition}, pp.\  12162--12171, 2024.

\bibitem[Zhang et~al.(2019)Zhang, Xu, Cuesta-Infante, and
  Veeramachaneni]{zhang2019robust}
Zhang, K.~A., Xu, L., Cuesta-Infante, A., and Veeramachaneni, K.
\newblock Robust invisible video watermarking with attention.
\newblock \emph{arXiv preprint arXiv:1909.01285}, 2019.

\bibitem[Zhang et~al.(2024)Zhang, Liu, Martin, Bearfield, Brun, and
  Guan]{zhang2024attack}
Zhang, L., Liu, X., Martin, A.~V., Bearfield, C.~X., Brun, Y., and Guan, H.
\newblock Attack-resilient image watermarking using stable diffusion.
\newblock \emph{Advances in Neural Information Processing Systems},
  37:\penalty0 38480--38507, 2024.

\bibitem[Zhang et~al.(2018)Zhang, Isola, Efros, Shechtman, and
  Wang]{zhang2018unreasonable}
Zhang, R., Isola, P., Efros, A.~A., Shechtman, E., and Wang, O.
\newblock The unreasonable effectiveness of deep features as a perceptual
  metric.
\newblock In \emph{Proceedings of the IEEE conference on computer vision and
  pattern recognition}, pp.\  586--595, 2018.

\end{thebibliography}
\bibliographystyle{icml2026}

\newpage
\appendix
\onecolumn

\section{Codebook Construction and Bit--Permutation Mapping}
\label{app:codebook}

\subsection{Permutation-based payload encoding}
\label{app:codebook:perm}

In ShapeMark, each group $g$ contains $s=4$ blocks,
denoted by an ordered tuple $(B_{1,g},B_{2,g},B_{3,g},B_{4,g})$.
A payload symbol $m_g\in\{0,1\}^k$ is embedded by applying a within-group block permutation
$\sigma\in S_4$:
\begin{equation}
\rho_{\sigma}\big(B_{1,g},B_{2,g},B_{3,g},B_{4,g}\big)
\;=\;
\big(B_{\sigma(1),g},B_{\sigma(2),g},B_{\sigma(3),g},B_{\sigma(4),g}\big).
\end{equation}
Decoding searches over a \emph{codebook} $\mathcal{C}\subseteq S_4$ and outputs the index of the best-matching codeword.

\subsection{Why $|S_4|=24$ but only $2^k=16$ codewords are used}
\label{app:codebook:why16}

Although there are $4!=24$ possible block permutations, the main setting encodes a fixed-length $k=4$ bits per group,
which naturally corresponds to a power-of-two codebook size $|\mathcal{C}|=2^k=16$.
Using all 24 permutations would either (i) require variable-length coding across groups to exploit the extra $\log_2(24)\approx 4.58$ bits,
or (ii) introduce a non-uniform many-to-one mapping from bit strings to permutations, complicating threshold calibration and potentially
inducing payload-dependent bias. We therefore intentionally select a 16-codeword subset, which can be interpreted as \emph{rate reduction}
to increase robustness under inversion noise and lossy distortions, analogous to using redundancy in channel coding.

In addition, we choose $\mathcal{C}$ to be \emph{balanced}: across the 16 codewords, each block index $i\in\{1,2,3,4\}$
appears equally often at each position $j\in\{1,2,3,4\}$, which prevents systematic position bias as the payload varies.

\subsection{Canonical codebook and deterministic mapping}
\label{app:codebook:table}

Table~\ref{tab:codebook} lists a canonical balanced codebook $\mathcal{C}_{\mathrm{base}}=\{\sigma^{(0)},\ldots,\sigma^{(15)}\}\subset S_4$.
We map the integer value $v\in\{0,\ldots,15\}$ (obtained from the 4-bit string) to a permutation by
\begin{equation}
\mathrm{Enc}_{\mathrm{base}}(v)=\sigma^{(v)}, \qquad
\mathrm{Dec}_{\mathrm{base}}(\sigma^{(v)})=v.
\end{equation}

\begin{table}[h]
\centering
\small
\caption{A canonical balanced permutation codebook for $s=4$ and $k=4$.
A permutation $\sigma=(a,b,c,d)$ means the group order becomes $(B_{a,g},B_{b,g},B_{c,g},B_{d,g})$.}
\label{tab:codebook}
\begin{tabular}{c c c}
\toprule
$v$ & 4-bit payload ($m_g$) & $\sigma^{(v)}$ (block order) \\
\midrule
0  & 0000 & (1,2,4,3) \\
1  & 0001 & (1,3,2,4) \\
2  & 0010 & (1,3,4,2) \\
3  & 0011 & (1,4,3,2) \\
4  & 0100 & (2,1,4,3) \\
5  & 0101 & (2,3,1,4) \\
6  & 0110 & (2,4,1,3) \\
7  & 0111 & (2,4,3,1) \\
8  & 1000 & (3,1,2,4) \\
9  & 1001 & (3,2,1,4) \\
10 & 1010 & (3,2,4,1) \\
11 & 1011 & (3,4,1,2) \\
12 & 1100 & (4,1,2,3) \\
13 & 1101 & (4,1,3,2) \\
14 & 1110 & (4,2,3,1) \\
15 & 1111 & (4,3,2,1) \\
\bottomrule
\end{tabular}
\end{table}

\section{Additional Evaluation Details}
\label{app:eval-details}

\subsection{Calibrating the detection threshold at $\mathrm{FPR}=10^{-6}$}
\label{app:fpr_calibration}

Our detector outputs a scalar score $S$ (Eq.~\eqref{eq:global-score}), where larger values indicate stronger evidence of a valid watermark under a claimed key $\kappa$. To operate at a target false positive rate $\alpha=10^{-6}$, we require a threshold $\tau_{\mathrm{det}}$ satisfying
\begin{equation}
\Pr\!\left(S \ge \tau_{\mathrm{det}} \mid H_0\right) \;=\; \alpha,
\label{eq:fpr_target}
\end{equation}
where $H_0$ denotes the null hypothesis that the queried image is \emph{not} watermarked under $\kappa$.

\paragraph{Challenge.}
With a finite null sample size $n$, the smallest non-zero empirical tail probability is $1/n$ (here $n=10{,}000$, i.e., $10^{-4}$), which is insufficient to directly estimate $\alpha=10^{-6}$ by counting. We therefore estimate $\tau_{\mathrm{det}}$ by \emph{tail extrapolation} from $n=10{,}000$ sampled null images.

\paragraph{Procedure (tail extrapolation from 10k null samples).}
We first collect null scores $\{S_i\}_{i=1}^{n}$ by running the full verification pipeline (inversion + decoding score computation) on $n=10{,}000$ null images and the claimed key $\kappa$. Let $\widehat{F}_0$ denote the (unknown) CDF of $S$ under $H_0$.

We choose a high threshold $u$ (e.g., the empirical $q$-quantile of $\{S_i\}$ with $q\in[0.95,0.99]$) and define exceedances
\begin{equation}
Y_i = S_i - u \quad \text{for all } i \text{ with } S_i > u.
\end{equation}
Let $n_u$ be the number of exceedances and $\hat{p}_u=n_u/n$ be the empirical tail probability at $u$.

We then fit a parametric model to the tail beyond $u$. In our implementation we use a peaks-over-threshold (POT) extrapolation, modeling $Y$ with a Generalized Pareto Distribution (GPD) with shape $\xi$ and scale $\beta$:
\begin{equation}
\Pr(Y > y) \approx \left(1+\xi \frac{y}{\beta}\right)^{-1/\xi}\quad (y\ge 0),
\end{equation}
which yields the tail approximation for $s\ge u$:
\begin{equation}
\Pr(S > s) \approx \hat{p}_u \left(1+\xi \frac{s-u}{\beta}\right)^{-1/\xi}.
\label{eq:gpd_tail}
\end{equation}
Finally, we solve $\Pr(S>\tau_{\mathrm{det}})=\alpha$ using Eq.~\eqref{eq:gpd_tail}, giving
\begin{equation}
\tau_{\mathrm{det}} \;=\; u \;+\; \frac{\beta}{\xi}\left[\left(\frac{\hat{p}_u}{\alpha}\right)^{\xi}-1\right]
\quad (\xi\neq 0),
\qquad
\tau_{\mathrm{det}} \;=\; u \;+\; \beta\log\!\left(\frac{\hat{p}_u}{\alpha}\right)
\quad (\xi=0).
\label{eq:tau_extrap}
\end{equation}
The special case $\xi=0$ corresponds to an exponential tail, which can be interpreted as a simple log-linear extrapolation of the survival function.

Using the above procedure with $n=10{,}000$ null samples, the resulting detection threshold is $\tau_{\mathrm{det}}=\mathbf{0.005542}$, which is used throughout all experiments to report TPR at $\mathrm{FPR}=10^{-6}$.

\paragraph{Remarks.}
This calibration produces an \emph{extrapolated} threshold from $10{,}000$ null samples and thus depends on the tail-model fit; it is used to report TPR at $\mathrm{FPR}=10^{-6}$ in a consistent manner across methods. For deployment-grade calibration at extremely low FPR, one should additionally increase the null sample size and/or validate the tail fit using larger held-out null sets and cross-key negatives.

\subsection{Details about the $t$-test}
\label{app:ttest}

For evaluating the CLIP score and FID, we generate $n_s=n_0=10$ image sets for a two-sample $t$-test.
We test the hypotheses
\begin{equation}
H_0:\ \mu_s=\mu_0, \qquad H_1:\ \mu_s\neq \mu_0,
\end{equation}
where $\mu_s$ and $\mu_0$ denote the mean metric values for watermarked and clean images, respectively.
The $t$-statistic is computed as
\begin{equation}
t = \frac{|\mu_s-\mu_0|}{S^\ast \sqrt{\frac{1}{n_s}+\frac{1}{n_0}}},
\qquad
S^\ast=\sqrt{\frac{(n_s-1)S_s^2+(n_0-1)S_0^2}{n_s+n_0-2}},
\end{equation}
where $S_s$ and $S_0$ are the sample standard deviations of the two groups.

\subsection{Detailed Results of Robustness Evaluation}
\label{app:full-results:perdist}

The main paper reports the attacked score averaged over nine distortions. For completeness, Table~\ref{tab:perdist_tpr_transposed} enumerate results under each distortion. The attacked average reported in the main paper is computed as
\begin{equation}
\mathrm{Metric}_{\mathrm{attacked}} = \frac{1}{9}\sum_{j=1}^{9}\mathrm{Metric}\big(\mathcal{A}_j(I)\big),
\end{equation}
where $\mathcal{A}_j$ denotes the $j$-th distortion operator.

\begin{table*}[ht]
\centering
\caption{Traceability (bit accuracy) for each individual distortion (clean + 9 attacks).}
\label{tab:perdist_tpr_transposed}
\setlength{\tabcolsep}{6pt} 
\begin{tabular}{lccccccccc}
\toprule
Distortion & dwtDct & dwtDctSvd & RivaGAN & StableSig & GS & PRCW & T2SMark & ShapeMark \\
\midrule
Clean     
& 0.8115 & 0.9909 & 0.9769 & 0.9912 & 1.000 & 0.9086 & 1.0000 & 1.000 \\
JPEG      
& 0.4953 & 0.5146 & 0.8037 & 0.7865 & 0.9876 & 0.7609 & 0.9945 & 0.9913 \\
Drop      
& 0.7613 & 0.7754 & 0.9422 & 0.9804 & 0.9651 & 0.7577 & 0.9989 & 0.9985 \\
Crop      
& 0.5950 & 0.6106 & 0.9298 & 0.9645 & 0.9492 & 0.7596 & 0.9975 & 0.9977 \\
Resize    
& 0.5097 & 0.8222 & 0.9388 & 0.5061 & 0.9936 & 0.7575 & 0.9994 & 0.9997 \\
G-Blur    
& 0.4967 & 0.6252 & 0.8117 & 0.4035 & 0.9796 & 0.7583 & 0.9954 & 0.9976 \\
M-Blur    
& 0.5189 & 0.9034 & 0.9462 & 0.6442 & 0.9967 & 0.7589 & 0.9996 & 0.9997 \\
G-Noise   
& 0.6174 & 0.7728 & 0.7894 & 0.7691 & 0.9406 & 0.7577 & 0.9535 & 0.9969 \\
S\&P      
& 0.5816 & 0.5089 & 0.7937 & 0.6951 & 0.9607 & 0.7584 & 0.9498 & 0.9320 \\
Bright    
& 0.5379 & 0.5158 & 0.7953 & 0.9061 & 0.9821 & 0.7854 & 0.9729 & 0.9870 \\
Adv.(ave)
& 0.5925 & 0.7040 & 0.8728 & 0.7647 & 0.9755 & 0.7763 & 0.9859 & 0.9870 \\
\bottomrule
\end{tabular}
\end{table*}

\section{Illustration of watermark embedding and extraction algorithms}

Algorithms~\ref{alg:SE2e} and~\ref{alg:detect2e} summarize the complete procedures for watermark embedding and extraction in ShapeMark. The algorithms consolidate the key steps introduced in Sections~\ref{sec:SE} and~\ref{sec:pdsr}, including separability-guided structural template construction, group-wise permutation-based payload encoding, payload-agnostic structural randomization, diffusion inversion, and codebook-based decoding. Presenting the methods in algorithmic form clarifies the order of operations, the role of the secret key, and the information flow between embedding and verification, enabling straightforward implementation and reproducibility.

\begin{algorithm}[t]
\caption{SE: Separable Index Template Construction and Payload-Conditioned Group-wise Block Permutation}
\label{alg:SE2e}
\begin{algorithmic}[1]

\STATE \textbf{Input:} identity key $\kappa$; payload $m$; number of bins $Q$ (default $4$); block size $b$; permutation codebook $\mathcal{C}\subseteq S_Q$ with encoder $\mathrm{Enc}$.

\STATE \textbf{Output:} encoded latent $z^{(e)}$; group template $\{\mathcal{G}_g\}$; ordered block list $\{\mathcal{I}_n\}$.

\STATE
\STATE \textbf{Step 1: Construct separable index template}

\STATE Sample latent $z \sim \mathcal{N}(0,I)$ using RNG seeded by $\kappa$.
\STATE Let $D=C\cdot H\cdot W$ and flatten $z$ into a vector of length $D$.
\STATE $\pi \leftarrow \mathrm{argsort}(|z|)$.
\STATE $\ell \leftarrow D/Q$.
\STATE $T \leftarrow \ell/b$.
\STATE $G \leftarrow T$.
\STATE $N \leftarrow D/b$.

\FOR{$q=1$ to $Q$}

\STATE $\mathcal{B}_q \leftarrow \{\pi((q-1)\ell+1),...,\pi(q\ell)\}$

\STATE Generate a key-reproducible random ordering of indices in $\mathcal{B}_q$ using $\kappa$.

\STATE Denote ordered list by $\tilde{\mathcal{B}}_q$.

\FOR{$t=1$ to $T$}

\STATE $\mathcal{I}_{q,t} \leftarrow (\tilde{\mathcal{B}}_q[(t-1)b+1],...,\tilde{\mathcal{B}}_q[tb])$

\STATE $n \leftarrow (q-1)T+t$

\STATE $\mathcal{I}_n \leftarrow \mathcal{I}_{q,t}$

\ENDFOR
\ENDFOR

\FOR{$g=1$ to $G$}

\STATE $\mathcal{G}_g \leftarrow (\mathcal{I}_{1,g},...,\mathcal{I}_{Q,g})$

\ENDFOR

\STATE
\STATE \textbf{Step 2: Payload-conditioned block permutation}

\STATE $k \leftarrow \log_2|\mathcal{C}|$

\STATE Split payload $m$ into $G$ chunks $m_1,...,m_G$.

\STATE Initialize $z^{(e)} \leftarrow z$.

\FOR{$g=1$ to $G$}

\STATE $\sigma_g \leftarrow \mathrm{Enc}(m_g)$

\FOR{$j=1$ to $Q$}

\STATE $z^{(e)}[\mathcal{I}_{j,g}] \leftarrow z[\mathcal{I}_{\sigma_g(j),g}]$

\ENDFOR
\ENDFOR

\STATE \textbf{Return:} $z^{(e)}$, $\{\mathcal{G}_g\}$, $\{\mathcal{I}_n\}$

\end{algorithmic}
\end{algorithm}

\begin{algorithm}[t]
\caption{Watermark Detection and Payload Decoding}
\label{alg:detect2e}
\begin{algorithmic}[1]

\STATE \textbf{Input:} image $I$; identity key $\kappa$; nonce $r$; parameters $(Q,b,s)$; codebook $\mathcal{C}$; decoder $\mathrm{Dec}$; inversion operator $\mathcal{D}^{\dagger}$; threshold $\tau_{\mathrm{det}}$.

\STATE \textbf{Output:} detection statistic $S$; decoded payload $\hat m$.

\STATE
\STATE \textbf{Step 1: Recover SE-aligned latent}

\STATE $\hat z^{(p)} \leftarrow \mathcal{D}^{\dagger}(I)$

\STATE Regenerate canonical latent $z \sim \mathcal{N}(0,I)$ using key $\kappa$.

\STATE Rebuild index template $\{\mathcal{G}_g\}$ and $\{\mathcal{I}_n\}$ using Algorithm~\ref{alg:SE2e} Step 1.

\STATE $\kappa_{\mathrm{pdsr}} \leftarrow \mathrm{KDF}(\kappa,r)$

\STATE $\tau \leftarrow \mathrm{PRP}_{\kappa_{\mathrm{pdsr}}}(N)$

\FOR{$n=1$ to $N$}

\STATE $\hat z^{(e)}[\mathcal{I}_n] \leftarrow \hat z^{(p)}[\mathcal{I}_{\tau^{-1}(n)}]$

\ENDFOR

\STATE
\STATE \textbf{Step 2: Codebook matching}

\STATE Initialize $\hat m \leftarrow \emptyset$

\STATE $S \leftarrow 0$

\FOR{$g=1$ to $G$}

\STATE $d_{\min} \leftarrow +\infty$

\STATE $d_{\mathrm{second}} \leftarrow +\infty$

\FOR{each $\sigma \in \mathcal{C}$}

\STATE compute

\[
d=\sum_{j=1}^{s}\|\hat z^{(e)}[\mathcal{I}_{j,g}]
-
z[\mathcal{I}_{\sigma(j),g}]\|_2^2
\]

\IF{$d<d_{\min}$}

\STATE $d_{\mathrm{second}} \leftarrow d_{\min}$

\STATE $d_{\min} \leftarrow d$

\STATE $\hat{\sigma}_g \leftarrow \sigma$

\ELSE

\IF{$d<d_{\mathrm{second}}$}

\STATE $d_{\mathrm{second}} \leftarrow d$

\ENDIF
\ENDIF

\ENDFOR

\STATE $\hat m_g \leftarrow \mathrm{Dec}(\hat{\sigma}_g)$

\STATE $\hat m \leftarrow \mathrm{Concat}(\hat m,\hat m_g)$

\STATE
\[
m_g =
\max\left(
0,
\frac{d_{\mathrm{second}}-d_{\min}}
{d_{\mathrm{second}}+\varepsilon}
\right)
\]

\STATE $S \leftarrow S+m_g$

\ENDFOR

\STATE $S \leftarrow S/G$

\STATE \textbf{Decision:} watermark present if $S \ge \tau_{\mathrm{det}}$

\STATE \textbf{Return:} $S$, $\hat m$

\end{algorithmic}
\end{algorithm}

\begin{figure*}[ht]
    \centering
    \includegraphics[
        width=\textwidth,
        trim=20 0 0 20,
        clip
    ]{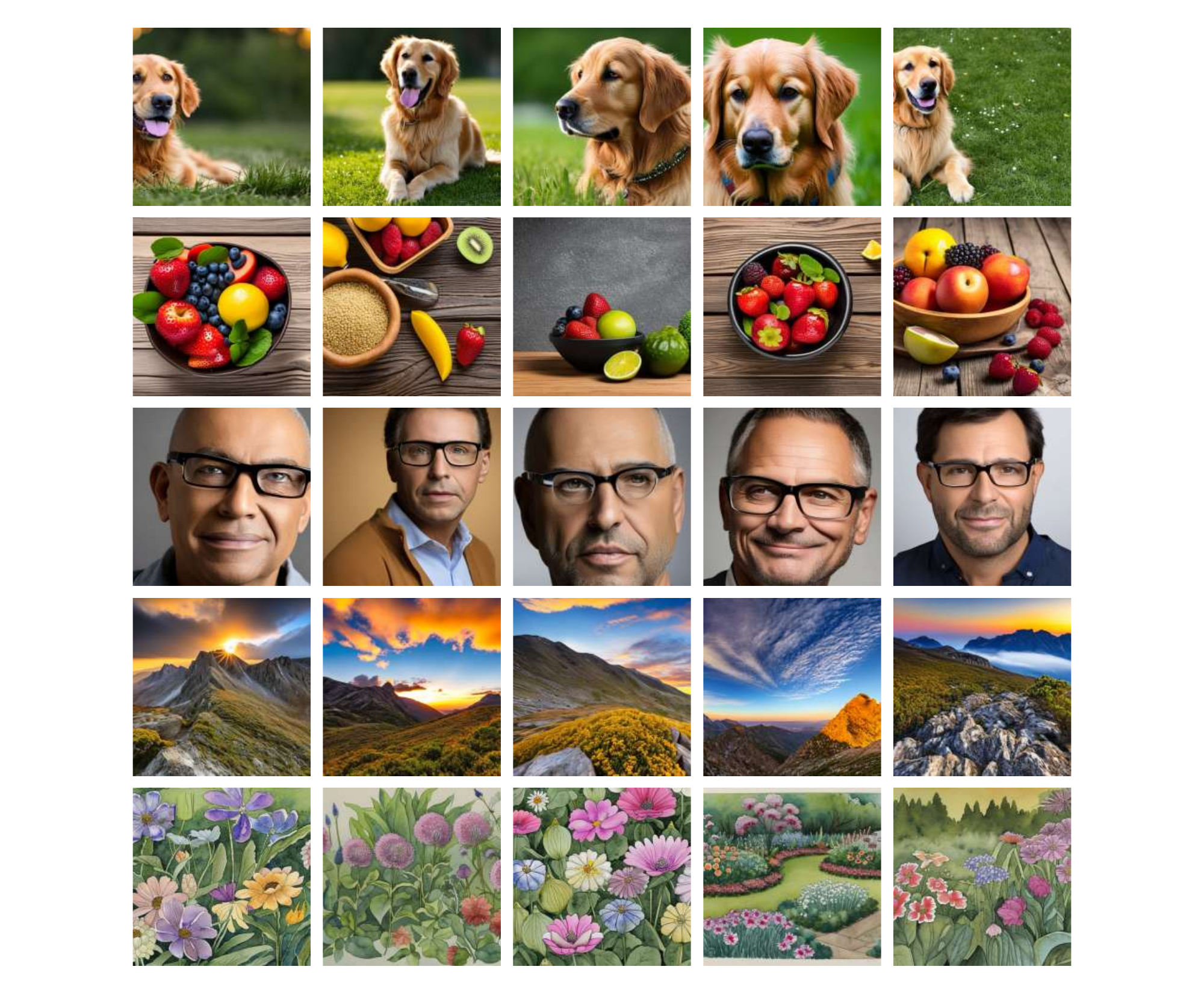}
    \caption{Qualitative examples of watermarked images generated from diverse text prompts.
The images are generated using the following prompts:
(1) “A close-up photo of a golden retriever dog sitting on grass, natural light, shallow depth of field”;
(2) “A bowl of fresh fruits on a wooden table, photorealistic, studio lighting”;
(3) “A head-and-shoulders portrait of a man wearing glasses, neutral expression, studio lighting”;
(4) “A wide-angle photo of a mountain landscape at sunrise, dramatic sky, high detail”; and
(5) “A watercolor illustration of flowers in a garden, soft colors”.
These prompts cover diverse semantic categories and visual styles, illustrating that our method preserves high perceptual quality while embedding watermarks across heterogeneous generation scenarios.}
    \label{appendix_img}
\end{figure*}


\end{document}